\documentclass[conference]{IEEEtran}
\ifCLASSINFOpdf
\else
\fi
\usepackage{graphicx}
\usepackage{graphics}
\usepackage{subfigure}
\usepackage{threeparttable}
\usepackage{float}
\usepackage{epsfig}
 \usepackage{epstopdf}
 \usepackage{amsmath}
\usepackage{color, soul}
\usepackage{soul}
\soulregister\cite7
\soulregister\citep7
\soulregister\citet7
\soulregister\ref7
\soulregister\pageref7
\begin{document}
%
\title{Probabilistic Two-Ray Model for Air-to-Air Channel in Built-Up Areas}
%
%
%
\author{\IEEEauthorblockN{Zhuangzhuang Cui\IEEEauthorrefmark{1}, Ke Guan\IEEEauthorrefmark{1}, C\'esar Briso\IEEEauthorrefmark{2}, Danping He\IEEEauthorrefmark{1}, Bo Ai\IEEEauthorrefmark{1} and Zhangdui Zhong\IEEEauthorrefmark{1}}
\IEEEauthorblockA{\IEEEauthorrefmark{1} State Key Lab of Rail Traffic Control and Safety, Beijing Jiaotong University, Beijing 100044, China \\
\IEEEauthorrefmark{2} ETSIS de Telecomunicaci\'on, Universidad Polit\'ecnica de Madrid, Madrid 28031, Spain \\
Corresponding Author: Ke Guan, kguan@bjtu.edu.cn\\}}

\maketitle

\begin{abstract}
In this paper, we present a probabilistic two-ray (PTR) path loss model for air-to-air (AA) propagation channel in built-up areas. Based on the statistical model of city deployment, the PTR path loss model can be applied to suburban, urban, dense urban, and high-rise urban. The path loss is optimally fitted as the Weibull distribution and its fluctuation is fitted as the Normal distribution in ray-tracing simulations. The good agreements between our model and ray tracing indicate the proposed model can provide a useful tool for accurate and quick prediction for aerial platforms. As an extended research of PTR model, we extract the shadowing factor by numerous simulations and propose the altitude-dependent shadowing model. The result shows that the proposed shadowing model has very good consistent with the measurement-based model, which indicates that our research performs well in the extensibility and generality.
\end{abstract}


%
\IEEEpeerreviewmaketitle

\section{Introduction}
\IEEEPARstart{H}{ow} altitude platforms (HAPs) could provide a possible alternative to the terrestrial and satellite provision of mobile services \cite{b1}. The aerial platforms can cooperatively work so as to accomplish the difficult task or realize seamless coverage for communication service such as unmanned aerial vehicle (UAV) swarm \cite{b2}. In this way, the robust and high-speed communication between platforms becomes very important. Therefore, it is critical to have a better understanding of the air-to-air (AA) propagation channel and provide an accurate and easy-to-use channel model.

The first concern should be taken into consideration is the altitude of the platform. Unfortunately, the uniform agreement on the specific altitude of HAP is not yet formed. However, a majority of studies are concentrated on few hundred meters with consideration of the safety and and the legal provisions. For example, ITUR M.2171 suggests that the low altitude unmanned aircraft (UA) should fly below 1500 m \cite{b3}. The Federal Aviation Administration (FAA) has issued rules for small UAVs with a maximum flight ceiling of 122 m above the ground level \cite{b4}. For sake of safety in built-up scenario, we focus on the altitude higher than the building in this paper so as to avoid collisions between platform and building. At such altitudes, the diffractions and high-order reflections can be neglected in the propagation because of the little probability. Moreover, even if there are diffractions and high-order reflections, the power of them is inappreciable because the large path length leads to severe propagation loss. Thus, the subject of this paper is based on the two-ray model which consists of the line-of-sight (LOS) and the first-order reflection. The results of a large number of measurement campaigns have indicate that the two-ray model can basically describe the channel behavior when there is no other scatterers around the transceiver, such as over-water scenario \cite{b5} and urban scenario when the platform is at high altitude \cite{b6}.

Facing many future applications in built-up areas, the free space path loss (FSPL) model is insufficient for the design and simulation of mobile systems provided by HAPs. The measurement campaign to clarify the channel characteristics of the AA propagation at C-band between small UAs was performed in \cite{b5}. The Rice channel model was extended to interpret multipath effects introduced by the flight altitude of UAVs on an IEEE 802.11 communication link based on measurements in \cite{b7}. Since channel measurements not only require high-precision instruments but also are restricted by scenes, there are limited measurement campaigns for AA channel. Thus, we focus on the analytical method based on the characteristics of scenario and the conditions of the propagation. Thanks to the high altitude of the platforms, the LOS path is not blocked so as to propagate in free space. However, the reflections could be from the roof or the ground. The critical issue of the two-ray model is the constructive and destructive superposition between the LOS and the reflection. Thus, modeling different types of reflections is an important concern in our paper.

The remainder of this paper is organized as follows. In Section II, we introduce the general city deployments defined by ITU-R and propose the probabilistic two-ray path loss model. In Section III, comprehensive validations by ray-tracing simulations and measurement-based for the proposed model are provided. Conclusions are drawn in Section IV.



\section{Probabilistic Two-Ray Path Loss Model}
\subsection{Scenario Description}
In order to set up a generic path loss model in built-up areas, the statistical ITU-R Rec. P.1410 model for building deployments is adopted \cite{b8}. An advantage of this model is that the area can be modeled without precise information concerning building shapes and distribution. The statistical model requires only three empirical parameters describing the built-up area: the ratio of the land area covered by buildings to the total land area ($\alpha$), the average number of buildings per unit area ($\beta$), and a parameter for determining the distribution of building height ($\gamma$). The values of three empirical parameters for different urban environments are listed in Table I. The probability density function (PDF) for the building height ($h_b$) based on the Rayleigh distribution is given by
\begin{equation}
p_r(h_b) =\frac{h_b}{\gamma}\exp(-\frac{h_b^2}{2\gamma^2}).
\end{equation}
\subsection{Proposed Model Description}
The FSPL model assumes that the energy of rays will neither be absorbed by obstacles nor be reflected or scattered when they propagate in free space. The calculation of the FSPL is given by
\begin{equation}
PL_{FS} [dB]=32.45+20\log_{10}(d)+20\log_{10}(f)
\end{equation}
where $d$ is the link distance in km, $f$ is the frequency in MHz.
\begin{table}[!htbp]
  \centering
  \caption{Parameters of Different Urban Environments According to The ITU-R Rec. P.1410 [6]} \label{table1}
\begin{tabular}{|c|c|c|c|}
  \hline
\textbf{Environment} & \textbf{$\alpha$} & \textbf{$\beta$} & \textbf{$\gamma$}\\
    \hline
Suburban & 0.1  & 750 & 8 \\
\hline
Urban & 0.3 & 500 & 15\\
    \hline
Dense Urban & 0.5 & 300 & 20 \\
\hline
High-rise Urban  & 0.5  & 300 & 50 \\
  \hline
\end{tabular}
\end{table}
The model only considers the LOS propagation in free space while neglects the reflections in environments, especially for built-up areas. Therefore, such a simple model is unsuitable in built-up areas because the reflected paths from the building and the ground exist for AA channel.

In this paper, we assume that the altitudes of HAPs represented by UAVs in figures are higher than the buildings because high altitudes are favorable for safe flights, in other words, could avoid the collision between the UAV and buildings, especially in the built-up scenarios. With such an assumption, the multipath components are composed of the direct path and the roof reflection or the ground reflection. Fig. 1(a) shows the three conditions of propagation for AA channel in the built-up environment: (a) the LOS and building reflection, (b) the LOS and ground reflection, and (c) the LOS only because the reflection is obstructed by buildings.
\begin{figure}[htbp]
\centering
\subfigure[]{\includegraphics[width=2.8in]{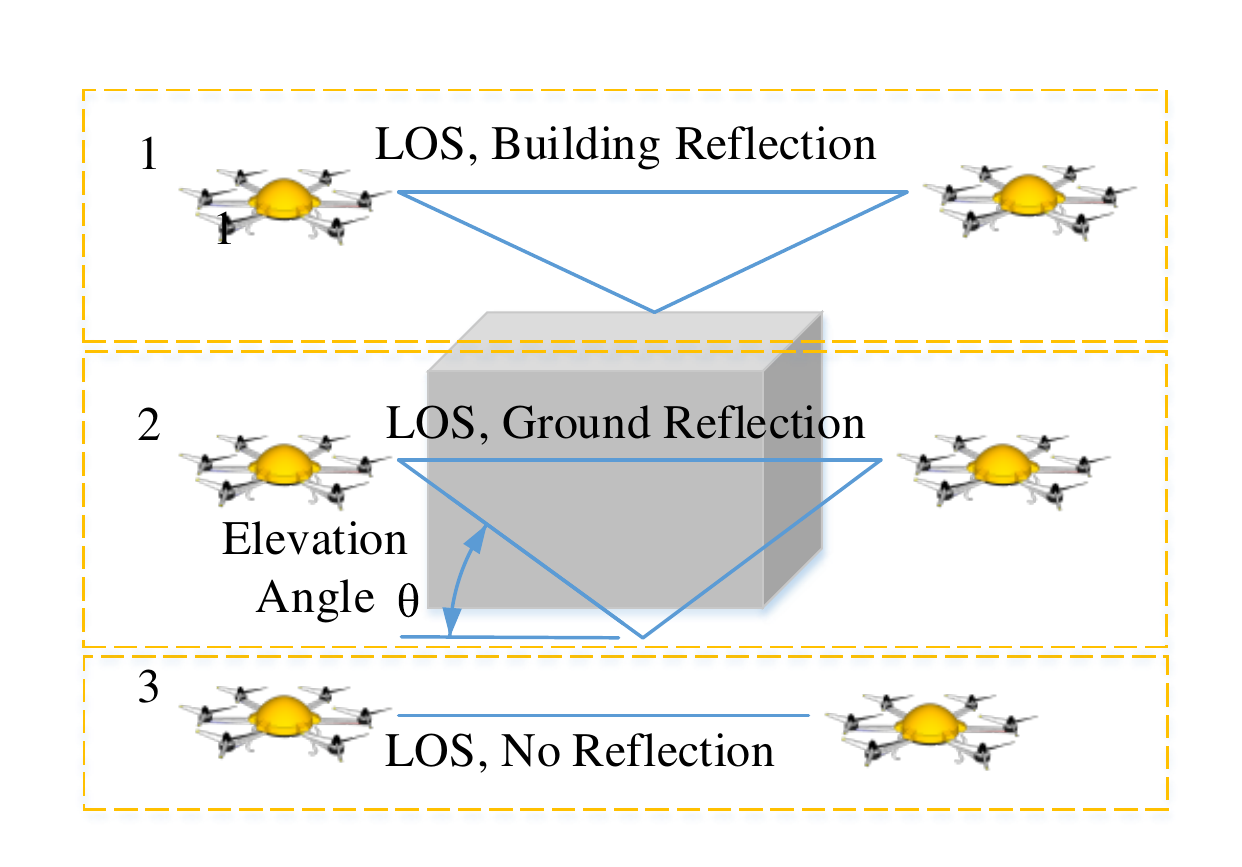}}   
\quad
\subfigure[]{\includegraphics[width=2.7in]{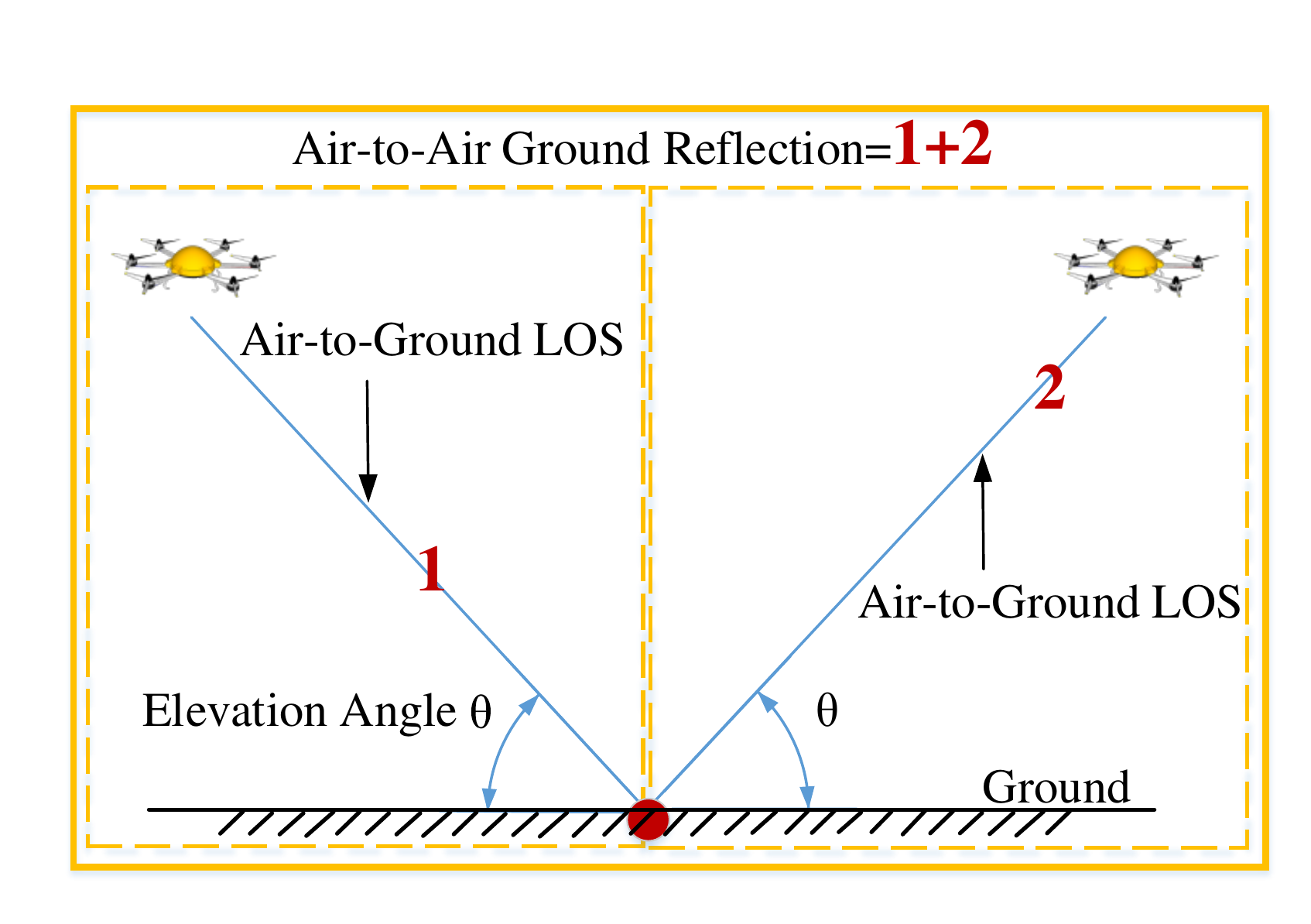}}
\caption{Propagation description: (a) Built-up areas with three conditions: (1) LOS and building reflection, (2) LOS and ground reflection, (3) LOS only; (b) Illustration of the AA channel ground reflection probability calculation.}
\label{fig_sim}
\end{figure}
According to the three conditions of propagation, the probabilistic two-ray (PTR) path loss model can be expressed as
\begin{equation}
\label{PL_PTR}
\begin{aligned}
PL_{PTR} [dB] = & 20 \cdot \log_{10}(4\pi\frac{d}{\lambda}|1+\Gamma_b\cdot \exp(i \Delta \varphi_b )\cdot \alpha \\&
+ \Gamma_g\cdot \exp(i \Delta \varphi_g)\cdot (1-\alpha)\cdot p_g|^{-1})
\end{aligned}
\end{equation}
where $\lambda$ is the wavelength, $\Gamma_b$ and $\Gamma_g$ are reflection coefficient for the building and ground, respectively. $\Delta\varphi_{b}$ and $\Delta\varphi_{g}$ are the phase difference between the building or ground reflection and LOS, respectively. In addition, $\alpha$ is the parameter in Table I, $p_g$ represents the probability of the ground reflection when the reflection point is on the ground. It must be emphasized that the ground reflection in the third channel condition in Fig.~1 is obstructed so that the AA channel only has the LOS path. The calculations of $\Delta\varphi_{b}$ and $\Delta\varphi_{g}$ are given by
\begin{equation}
\label{delta_fi}
 \Delta\varphi_{b/g}=\frac{2\pi}{\lambda}(d_{LOS}-d_{REF,b/g})
\end{equation}
where $d_{REF,b}$ and $d_{REF,g}$ are the path length of building reflection and ground reflection, respectively. $d_{REF,b}=\sqrt{d^2+4(h-h_b)^2}$ and $d_{REF,g}=\sqrt{d^2+4h^2}$. $h$ is the altitude of HAP and $d$ is the horizontal distance between two HAPs.

In Eq. (3), the reflection coefficient $\Gamma_{b/g}$ can be obtained by $\frac{\sin\theta-\sqrt{(\varepsilon_r+\cos^2\theta)}}{\sin\theta+\sqrt{(\varepsilon_r+\cos^2\theta)}}$ for horizontal polarization (HP) and $\frac{\sin\theta-\sqrt{(\varepsilon_r+\frac{\cos^2\theta}{\varepsilon_r})}}{\sin\theta+\sqrt{(\varepsilon_r+\frac{\cos^2\theta}{\varepsilon_r})}}$ for vertical polarization (VP) where $\theta$ is the elevation angle (as shown in Fig. 1) and $\varepsilon_r$ is the real part of relative permittivity.

\subsection{Ground Reflection Probability}
A challenging problem is to obtain the ground reflection probability, which is rather related to the building deployments. However, the LOS probability for AG channel has been proposed and validated in \cite{b9}. In order to conform to the Snell's law for the ground reflection in the AA channel, thus, the two direct paths in AG channels have the same elevation angle. Moreover, the LOS probability $p_{LOS}^{AG}$ is only defined as the function of the elevation angle $\theta$. It is noticed that only when the path 1 and 2 (see Fig. 1(b)) exist at the same time, the ground reflection of the AA channel exists. Moreover, the same deployment model from ITU-R is used in our paper, we can refer to the conclusion of LOS probability of AG channel in \cite{b9}. Thus, the ground reflection (GR) probability of AA channel can be developed from AG channel and is given by
\begin{equation}
\label{Ground_Reflection}
p_{GR}^{AA}(\theta) =(p_{LOS}^{AG} (\theta))^2 = (a-\frac{a-b}{1+(\frac{\theta-c}{d})^e})^2
\end{equation}
where $a$, $b$, $c$, $d$, $e$ are the empirical parameters given in Table II for the four typical environments. Note that the geometrical ground reflection is independent of the system frequency, also that (5) is generic and can be used for any elevation angle.
\begin{table}[!htbp]
  \centering
  \caption{Parameters for Ground Reflection Probability \cite{b10}} \label{table2}
  \begin{tabular}{|c|c|c|c|c|c|}
  \hline
\textbf{Environment} & \textbf{$a$} & \textbf{$b$} & \textbf{$c$}  & \textbf{$d$} & \textbf{$e$} \\
    \hline
Suburban & 101.6  & 0 & 0 & 3.25 & 1.241\\
\hline
Urban & 120.0 & 0 & 0 &  24.30 & 1.229 \\
    \hline
Dense Urban & 187.3  & 0 & 0 &  82.10 & 1.478 \\
\hline
High-rise Urban  & 352.0  & -1.37 & -53  &  173.80 & 4.670 \\
  \hline
\end{tabular}
\end{table}

The ground reflection probabilities with respect to elevation angles in different environments are depicted in Fig. 2. The probability is the mean value of the data obtained for all azimuth angles by means of ray-tracing simulations \cite{b9}. First, buildings are randomly generated using the statistical parameters in Table I. Then, the position of the HAP is determined for each point on the streets with a fine grid for a given elevation and azimuth angles in order to simulate and analyze an extremely large number of scenarios \cite{b11}. Note that the ground reflection probability is independent of the azimuth angle and only related to the elevation angle. The result in Fig. 2 indicates that the sparser building in the scenario, the higher ground reflection probability, which is in line with the realistic situation that the larger area of land leads to the relatively higher ground reflection probability.

\section{Validations for Proposed Model and Numerical Results}
\subsection{Validation for Ground Reflection Probability}
For the practicability of the proposed model, necessary validations are provided herein. First, the ground reflection probability is compared with \cite{b9} where for a 1.4~km$\times$1.4~km area of central Bristol, the number of buildings $N=717$, and the building coverage area $S_b=0.552$ km$^2$. Thus, $\alpha = 0.28$, $\beta=365$ buildings/km$^2$. The mean LOS probability as a function of elevation angle is obtained by ray-tracing simulation with 350,000 locations with a grid spacing of 2~m. We get the probability of ground reflection (the square of LOS probability in AG channel) in such environment. As shown in Fig. 3, similar results indicate that our model is accurate and practicable. The reason that the ground reflection probability in \cite{b9} slightly larger than our result in $15^\circ$-$65^\circ$ is that the building coverage ($\alpha$) and number ($\beta$) in Bristol city are smaller than the regular urban where $\alpha=0.3$ and $\beta=500$.
\begin{figure}[thbp]
\centering
\includegraphics[width=2.5in]{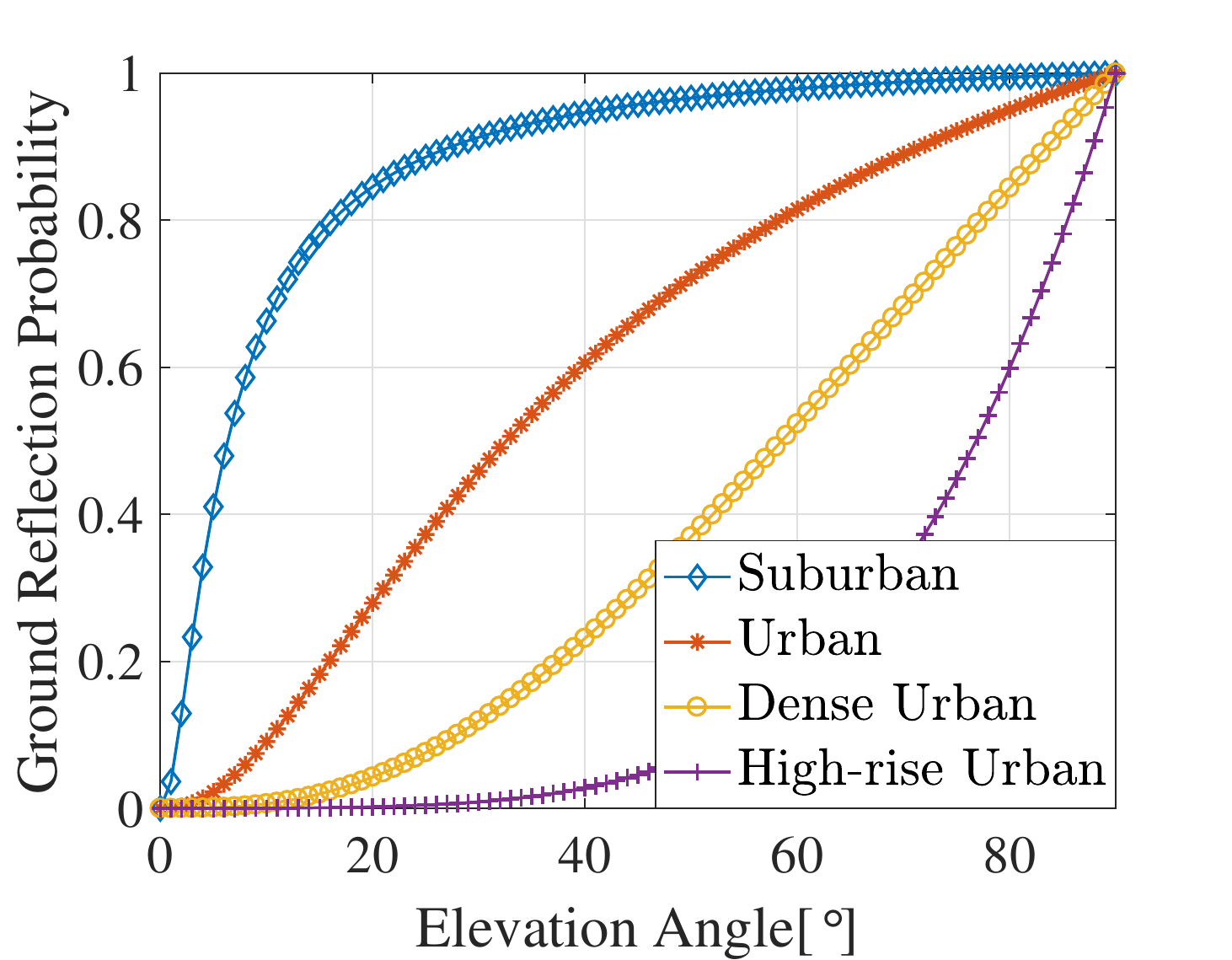}
\caption{Ground reflection probability vs. elevation angle for four environments.}
\label{fig_sim}
\end{figure}
\begin{figure}[htbp]
\centering
\includegraphics[width=2.5in]{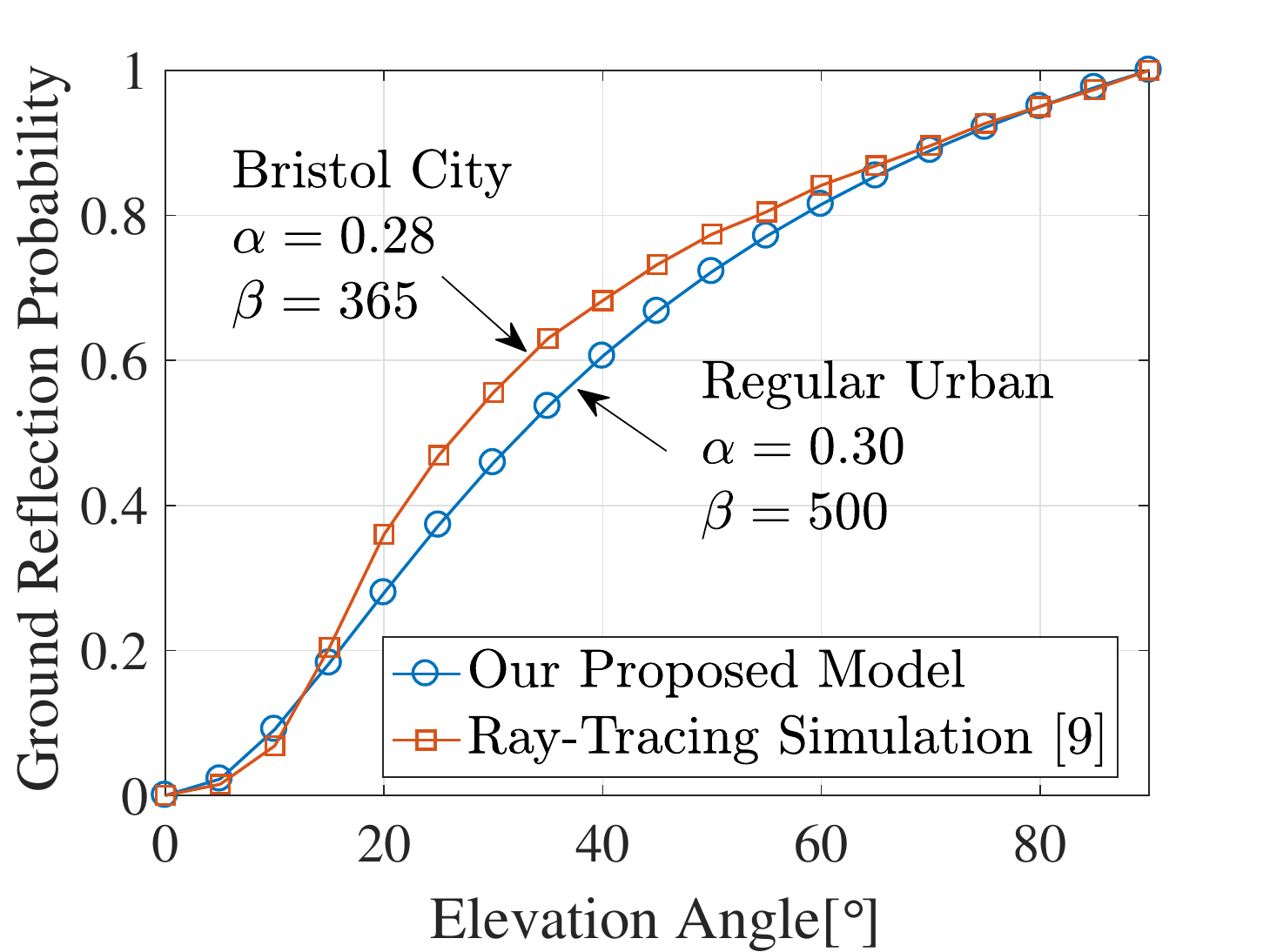}
\caption{Validation by the ray-tracing simulations from \cite{b9} in urban scenario.}
\label{fig_sim}
\end{figure}
\subsection{Ray-Tracing Simulations}
Ray-tracing simulations used to compare with our proposed model are conducted in the urban scenario. As shown in Fig.~4, the top view and three-dimensional SketchUp model with the size of (472~m $\times$ 472~m) is selected. In this study, it has been selected to shape the virtual-city environment similar to Manhattan grid \cite{b9}; an array of structures of an assumed square plot of width $W$, and inter-building spacing of $S$ where $W$ and $S$ are measured in meters, and can be linked to the ITU-R statistical parameters, since by definition: $\alpha$ = $(N_bW^2)$/$(1000D)^2$, where $D$ is the map side measured in kilometers (for a square patch), and $N_b$ is the number of building. On the other hand $\beta=N_b/D^2$. Thus, we can calculate $W$ and $S$ as the following: $W=1000\sqrt{\alpha/\beta}$ and $S=1000/\sqrt{\beta}-W$. In our simulation, $N_b=121$, $D=0.472$~km, $W=24.49$~m, and $S=20.23$~m. Note that the height of buildings are randomly generated and follows the Rayleigh distribution with $\gamma=15$ and the $\varepsilon_r$ of building and ground is set to 4.44 and 3, respectively. Note that the same parameters are used in theoretical simulations for our model.
\begin{figure}[htbp]
\centering
\subfigure[]{\includegraphics[width=2.5in]{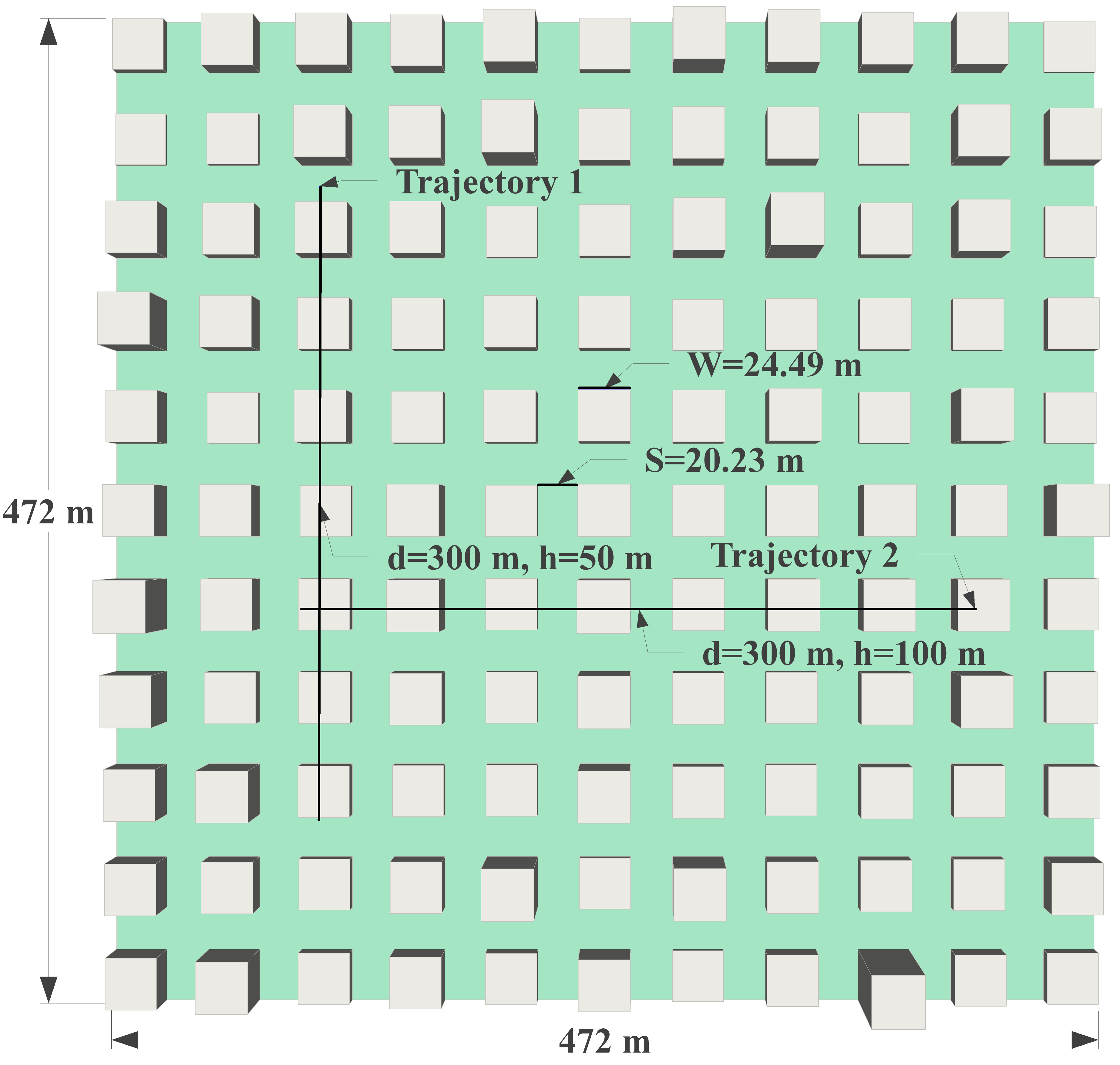}}
\quad
\subfigure[]{\includegraphics[width=3in]{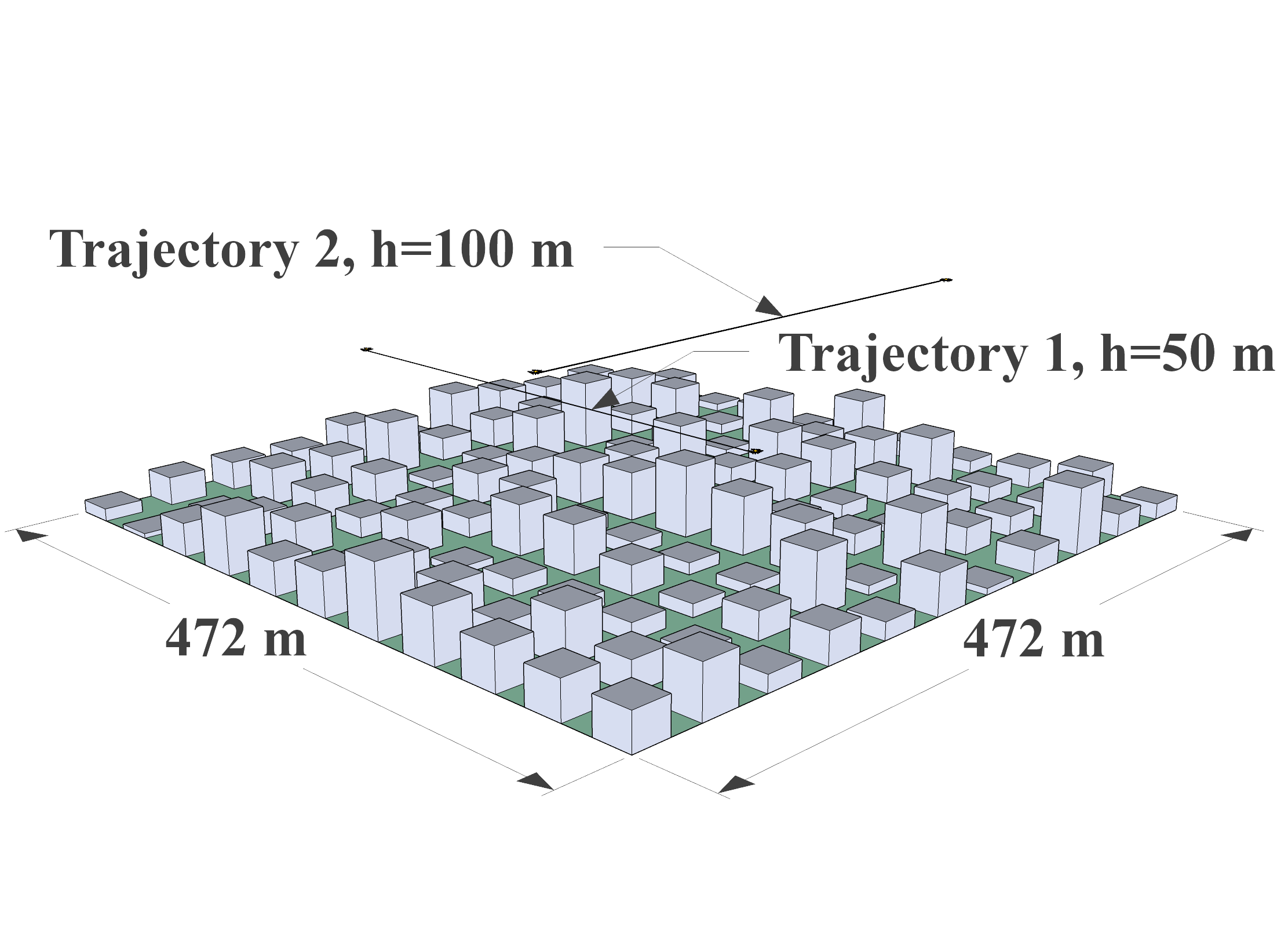}}
\caption{Regular urban: (a) Top view; (b) Three-dimensional SketchUp model.}
\label{fig_sim}
\end{figure}

Since part of the L-band is currently being proposed for the control link and the C-band for the payload for most communication scenarios \cite{b2}, the center frequency is selected at 4 GHz. As shown in Fig. 4, we select two trajectories that one is at 50 m and the other is at 100 m with a distance from 0 to 300 m between two HAPs. The results of path loss for ray tracing and our proposed model are illustrated in Fig. 4. The path loss exponents (PLEs) are all close to 2 for different methods of modeling regardless of the the altitude of the platform. Although the path loss results for our model and ray-tracing simulation have similar trends with the free space model, the FSPL is incapable of describing the channel behaviour, especially for fluctuations which caused by the reflections. Besides, for the aspect of the design of communication system, it is also significant to take the reflections into consideration because the reflections could lead to the large root-mean-square delay spread which is a significant parameter for cancelling the inter-symbol interference (ISI). Thus, we further compare the path loss fluctuations which also can be termed as shadow fading in Fig. 6(b). The results show that our model is consistent with the ray-tracing simulations.

\begin{figure}[htbp]
\centering
\subfigure[]{\includegraphics[width=2.5in]{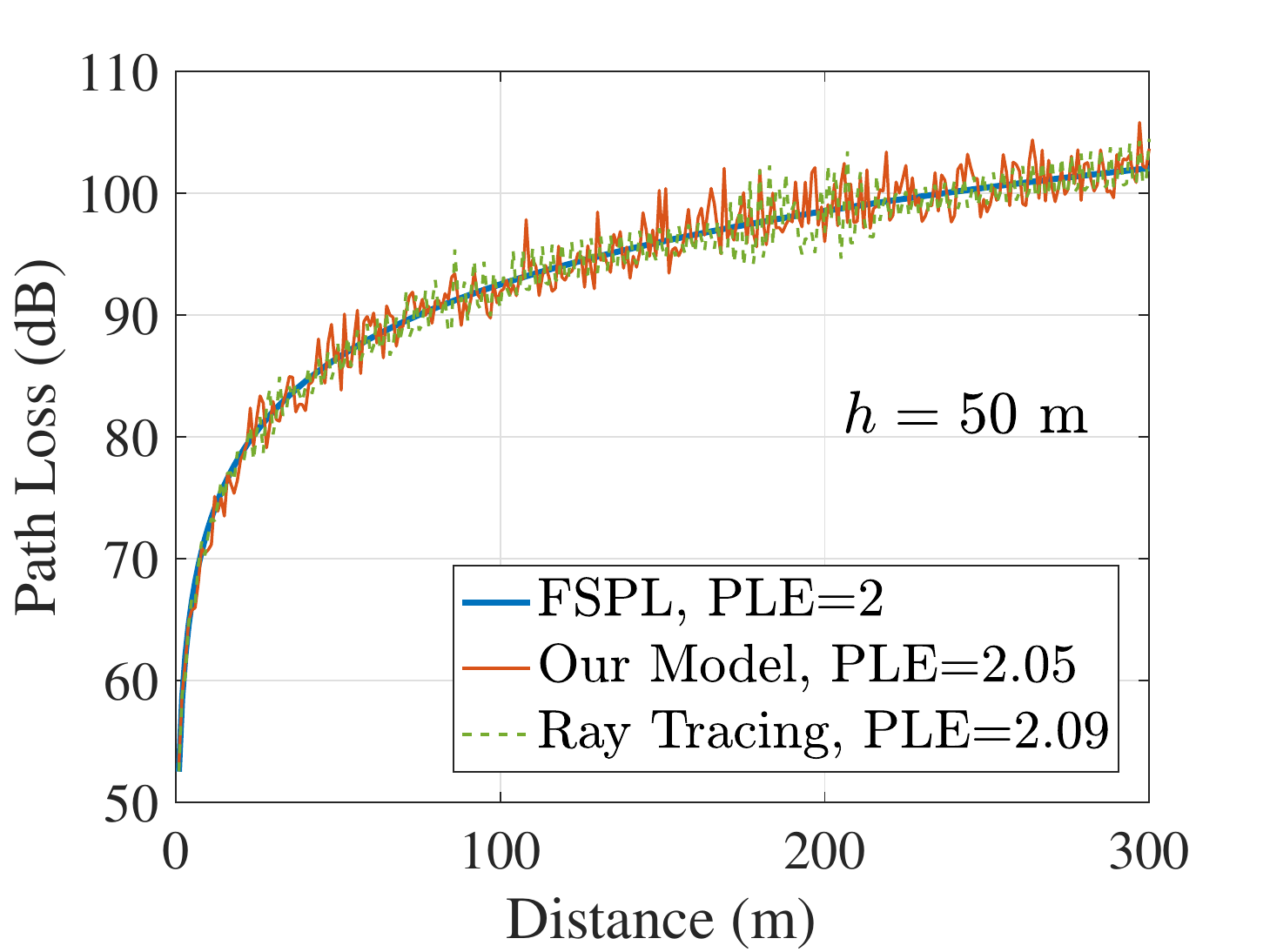}}
\quad
\subfigure[]{\includegraphics[width=2.5in]{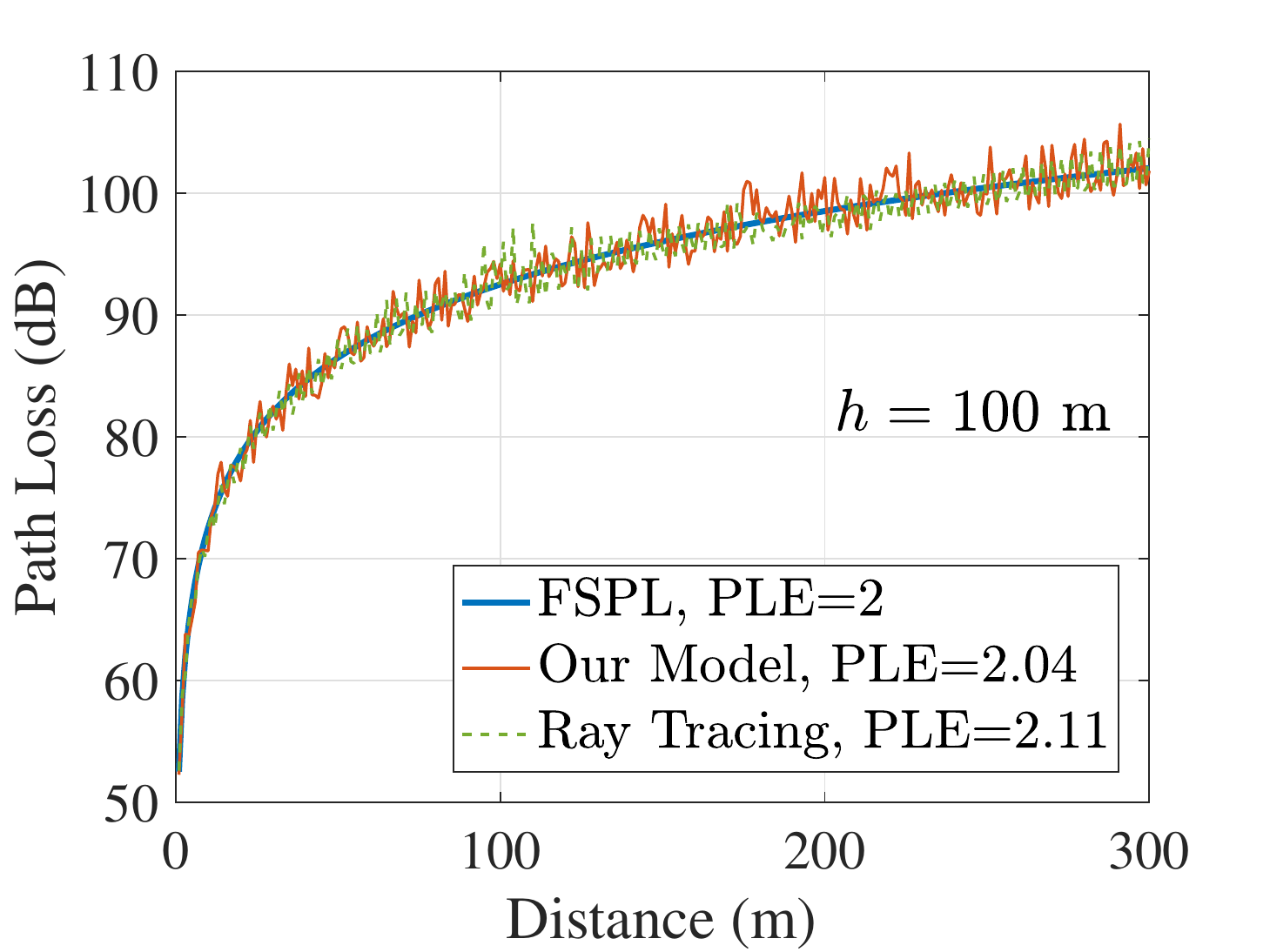}}
\caption{Path loss results for the FSPL, our proposed model, and the ray-tracing simulation where (a) $h$ = 50 m and (b) $h$ = 100 m. }
\end{figure}

\begin{figure}[htbp]
\centering
\subfigure[]{\includegraphics[width=2.5in]{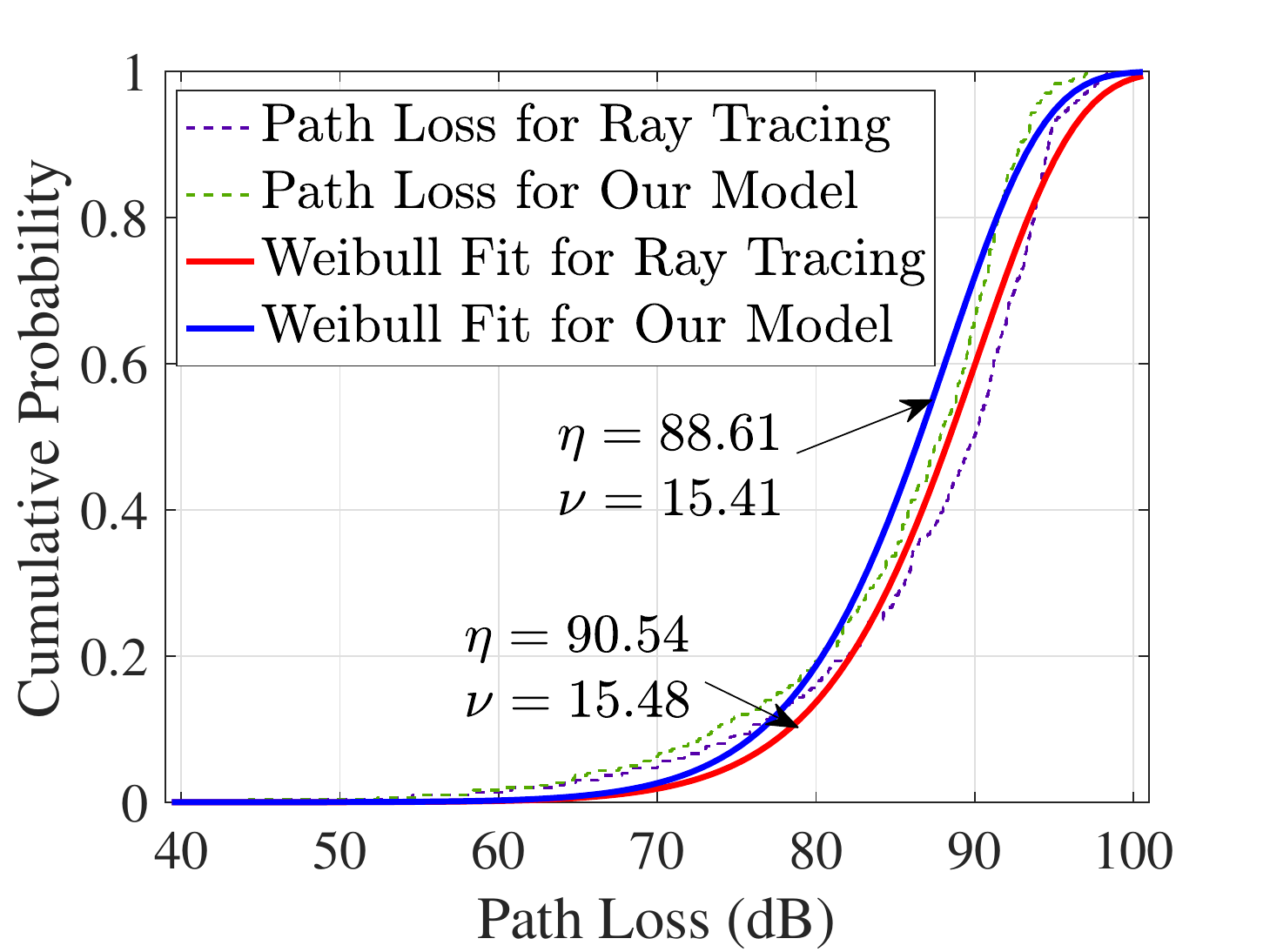}}
\quad
\subfigure[]{\includegraphics[width=2.5in]{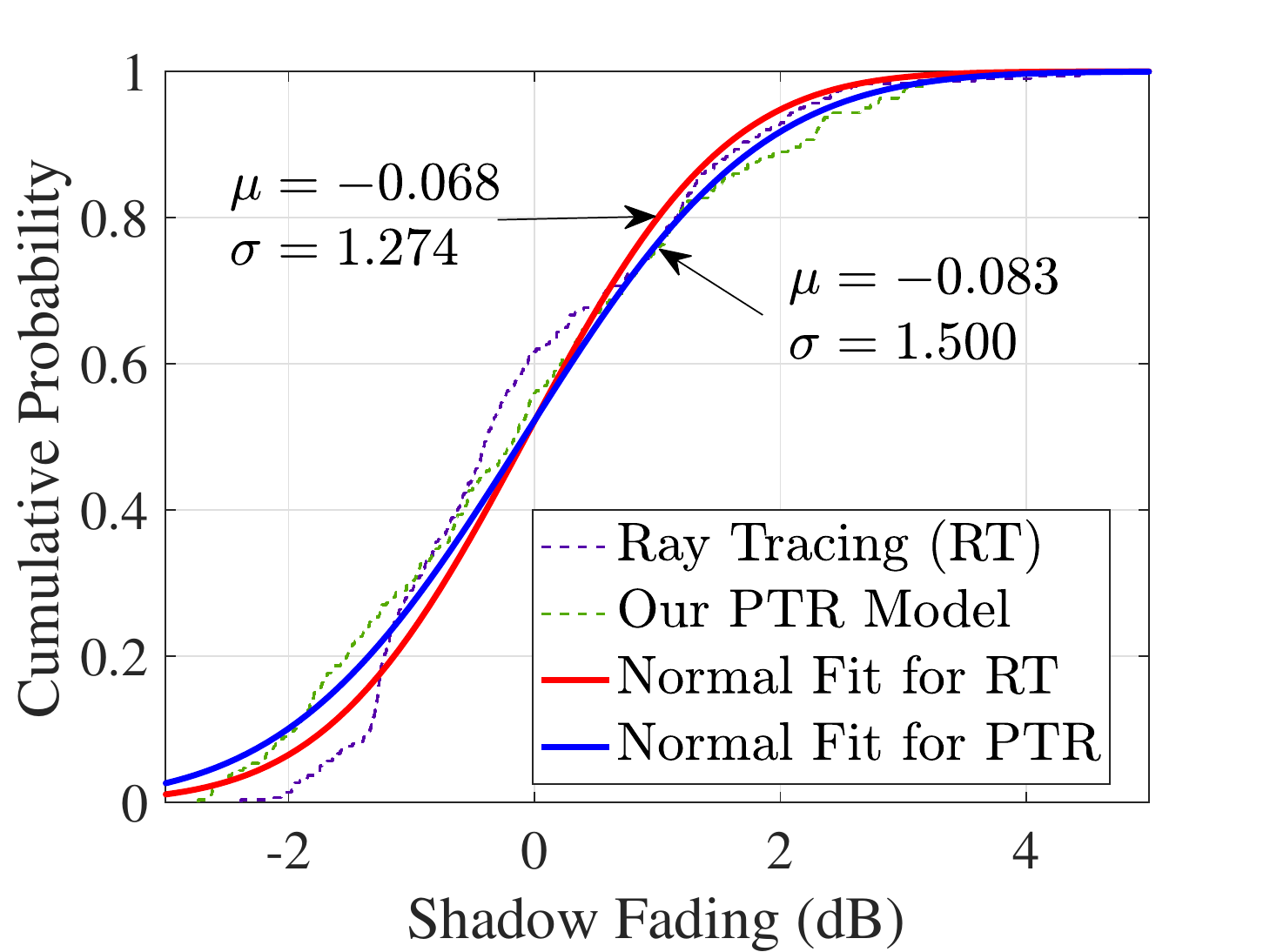}}
\caption{CDFs and fits (when $h$ = 100 m): (a) Path loss; (b) Shadow fading. }
\end{figure}

The cumulative probability distribution functions (CDFs) of path loss and shadow fading are fitted in Fig. 6 for ray tracing and our model, respectively. Through compared with many distributions such as Rician, Rayreigh, etc., the result is found that the path loss follows the Weibull distribution given by
\begin{equation}
\label{Weibull}
f(x,\eta, \nu) = \frac{\nu}{\eta}(\frac{x}{\eta})\exp({-(x/\eta)^\nu})
\end{equation}
where $x$ represents the path loss. Through the fitting, $\eta=90.54$ and $\nu =15.48$ for ray tracing. In our model, $\eta=88.61$ and $\nu =15.41$. The difference of the shape parameter ($\nu$) between our model and the ray tracing is 0.07, which indicates that our model for AA propagation channel could provide accurate prediction as the deterministic channel modeling such as ray tracing. In addition, the shadow fading is fitted as Normal distribution which is given by
\begin{equation}
\label{Normal}
f(y,\mu, \sigma) = \frac{1}{\sigma\sqrt{(2\pi)}}\exp({-\frac{(x-\mu)^2}{2\sigma^2}})
\end{equation}
where $y$ represents the value of shadow fading which actually is the path loss fluctuation because the LOS path is not blocked. By Normal fitting, $\mu=-0.068$ and $\sigma =1.274$ for ray tracing. In our model, $\mu=-0.083$ and $\sigma =1.500$. Similar results further indicate the accuracy of the proposed model.
\begin{figure}[thbp]
\centering
\subfigure[]{\includegraphics[width=1.65in]{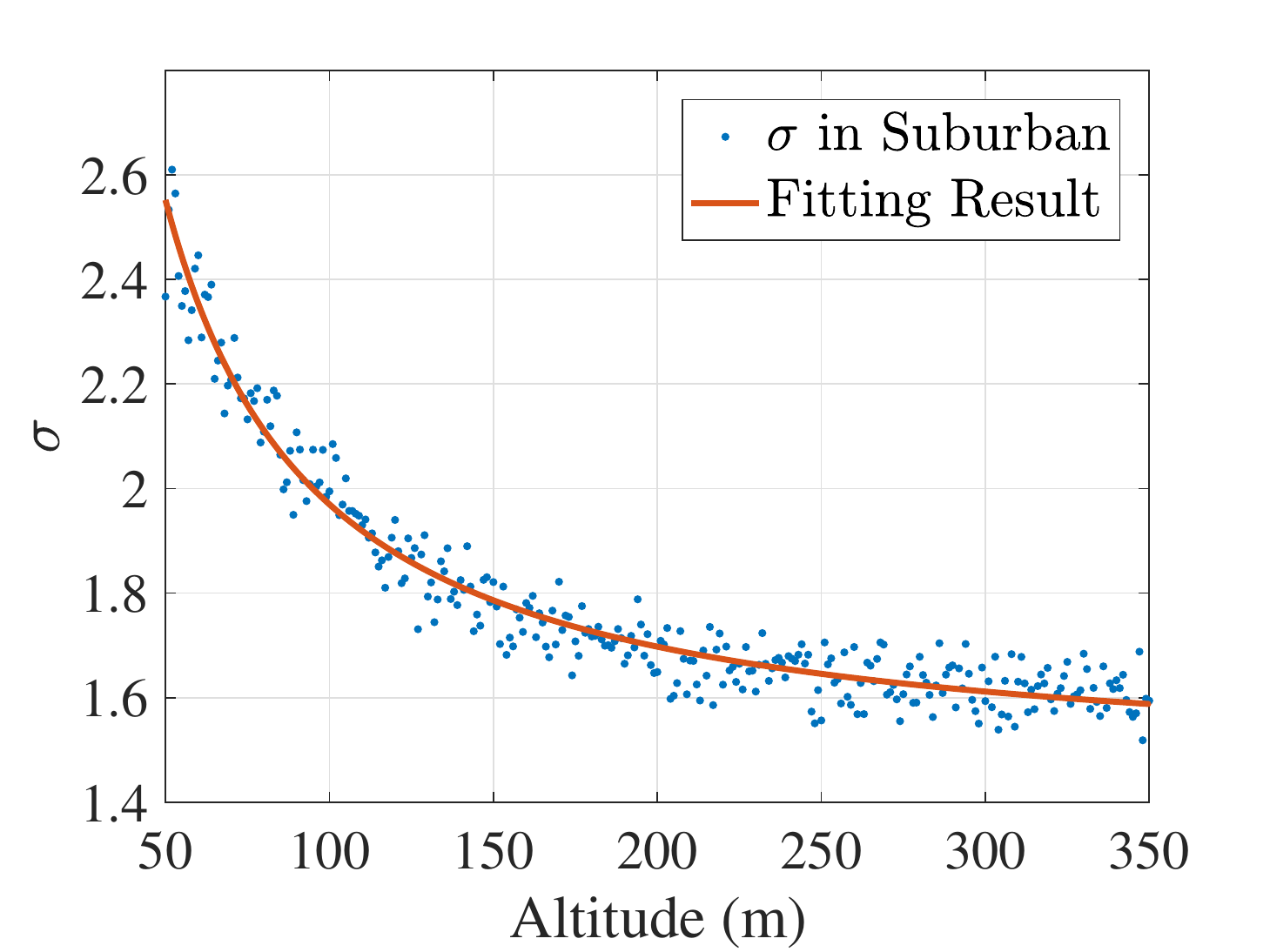}}
\quad
\subfigure[]{\includegraphics[width=1.65in]{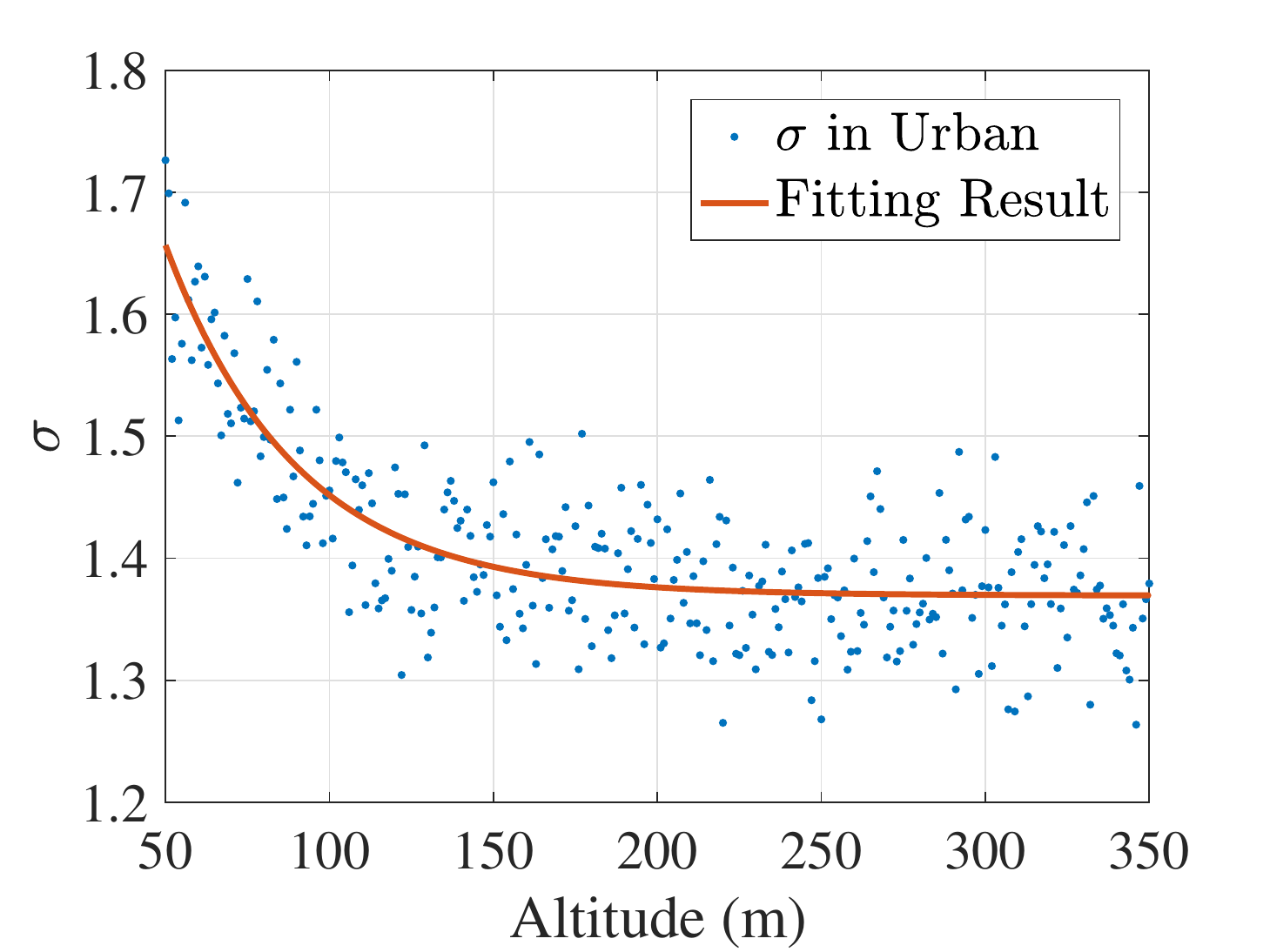}}
\quad
\subfigure[]{\includegraphics[width=1.65in]{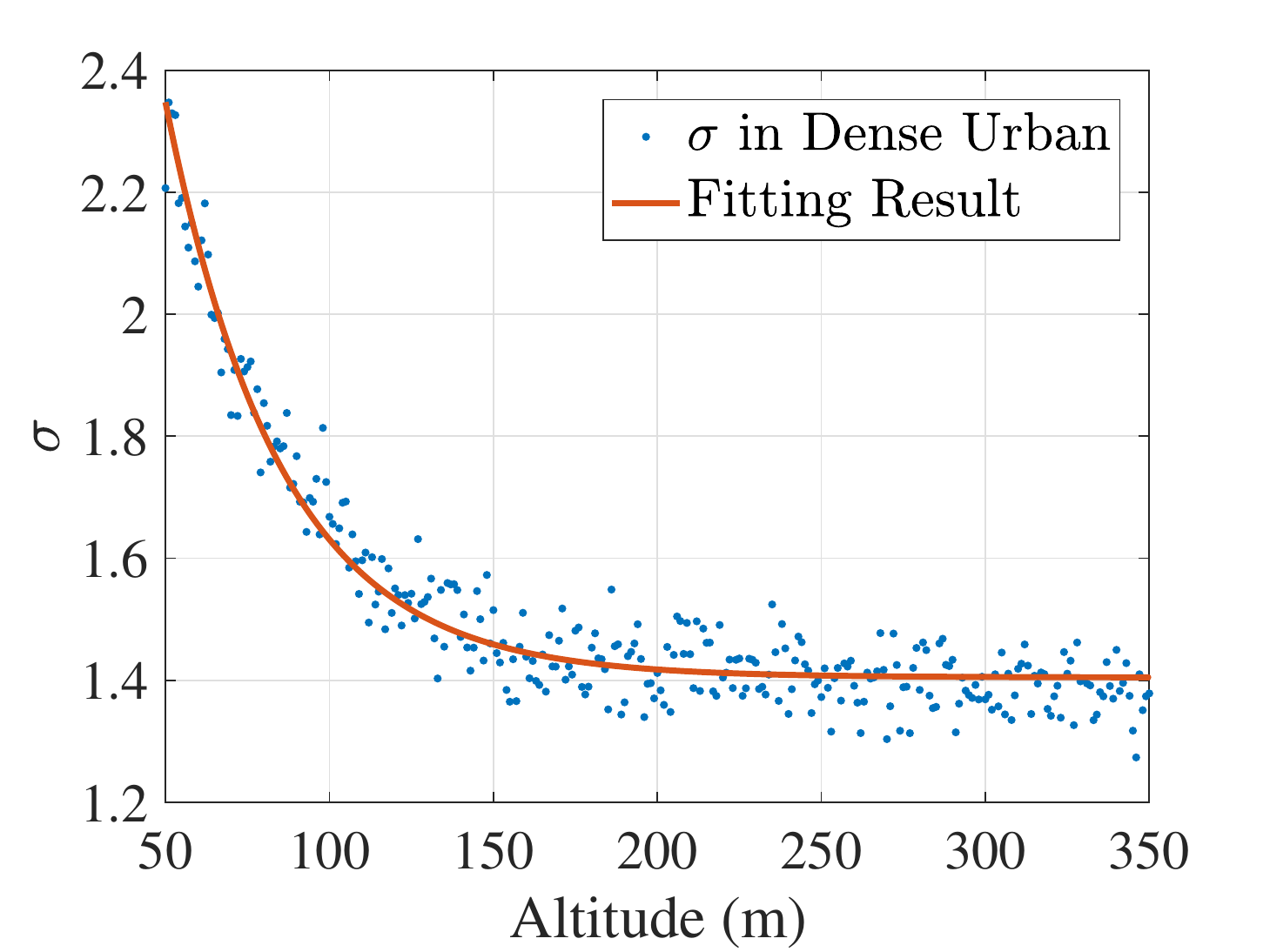}}
\quad
\subfigure[]{\includegraphics[width=1.65in]{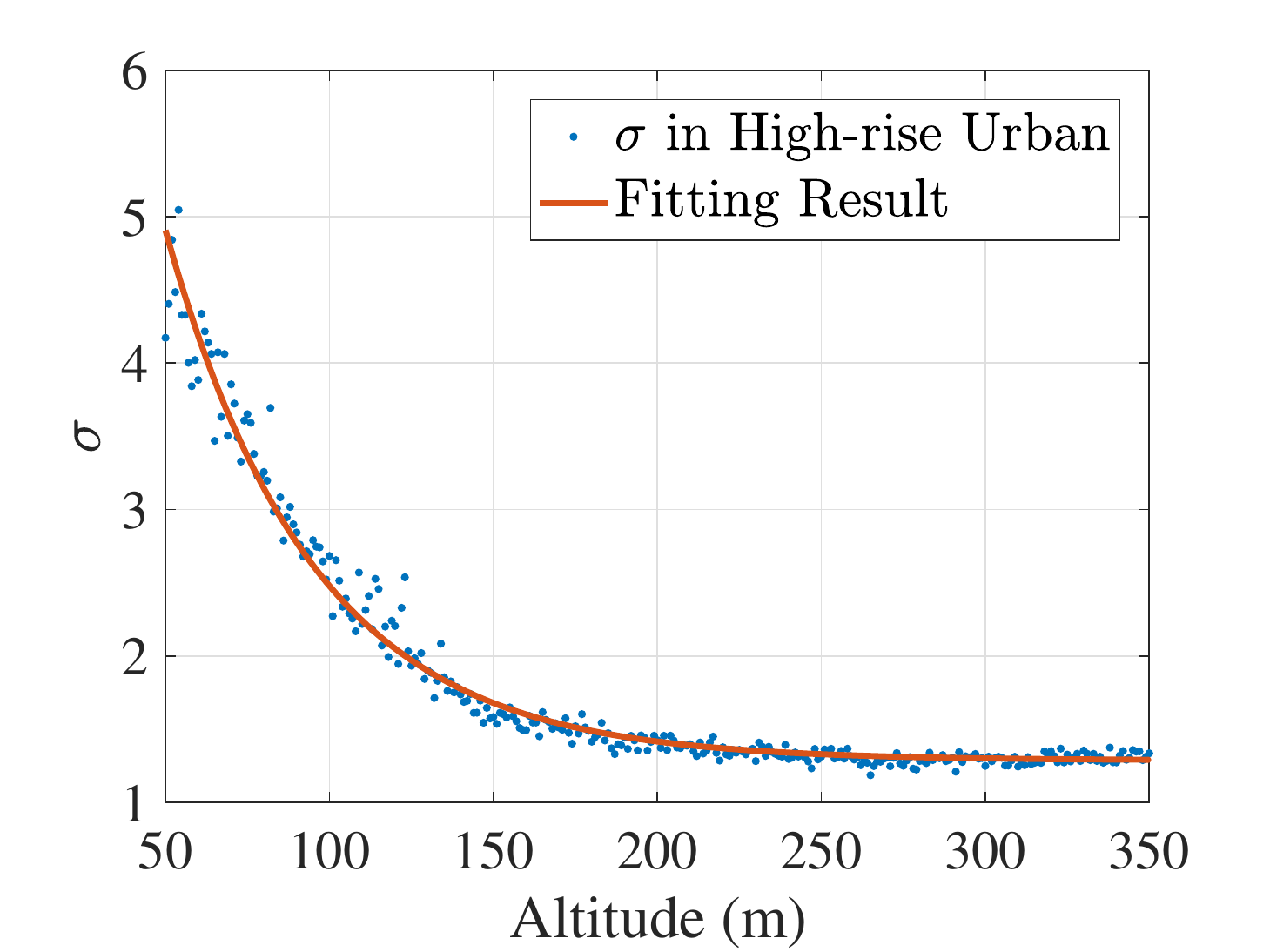}}
\caption{Shadowing factor $\sigma$ with respect to the altitude in different environments: (a) Suburban; (b) Urban; (c) Dense Urban; and (d) High-rise Urban. }
\end{figure}
\subsection{Altitude-Dependent Shadowing Factor Model}
An obvious phenomenon in Fig. 5 is that as the height rises, the fluctuation of path loss becomes small. Therefore, it is meaningful to establish an altitude-dependent shadowing factor model. Such a model in \cite{b7} is presented and verified by AA channel measurements. The measurement-based model is expressed as
\begin{equation}
\label{sigma_h in [5]}
\sigma [dB] = p_1\cdot h^{-q_1}+r_1
\end{equation}
where $p_1$, $q_1$ and $r_1$ are parameters that related to the characteristics of environments. Based on our proposed model, we simulated the path loss at various altitudes in different environments and thereby obtain the shadowing factors shown in Fig. 7. We found that the shadowing factors in different environments can be optimally fitted as the same form which is expressed as
\begin{equation}
\label{sigma_h}
\sigma [dB] = p_2\cdot \exp(-q_2\cdot h)+r_2
\end{equation}
where $p_2$, $q_2$ and $r_2$ are also environmentally relevant. In order to validate our proposed altitude-dependent shadowing model, we compare our model and the model in [7] based on measurement in suburban and the results are shown in Fig. 8. The result shows a good agreement between our model and the measurement-based model. By using the statistical parameters of built-up areas, the results of the altitude-dependent shadowing factor model ($p_2$, $q_2$ and $r_2$) for various environments can be easily obtained and listed in Table III. The results can be used for the simulation and optimization of the physical layer.

\begin{figure}[htbp]
\centering
\includegraphics[width=2.5in]{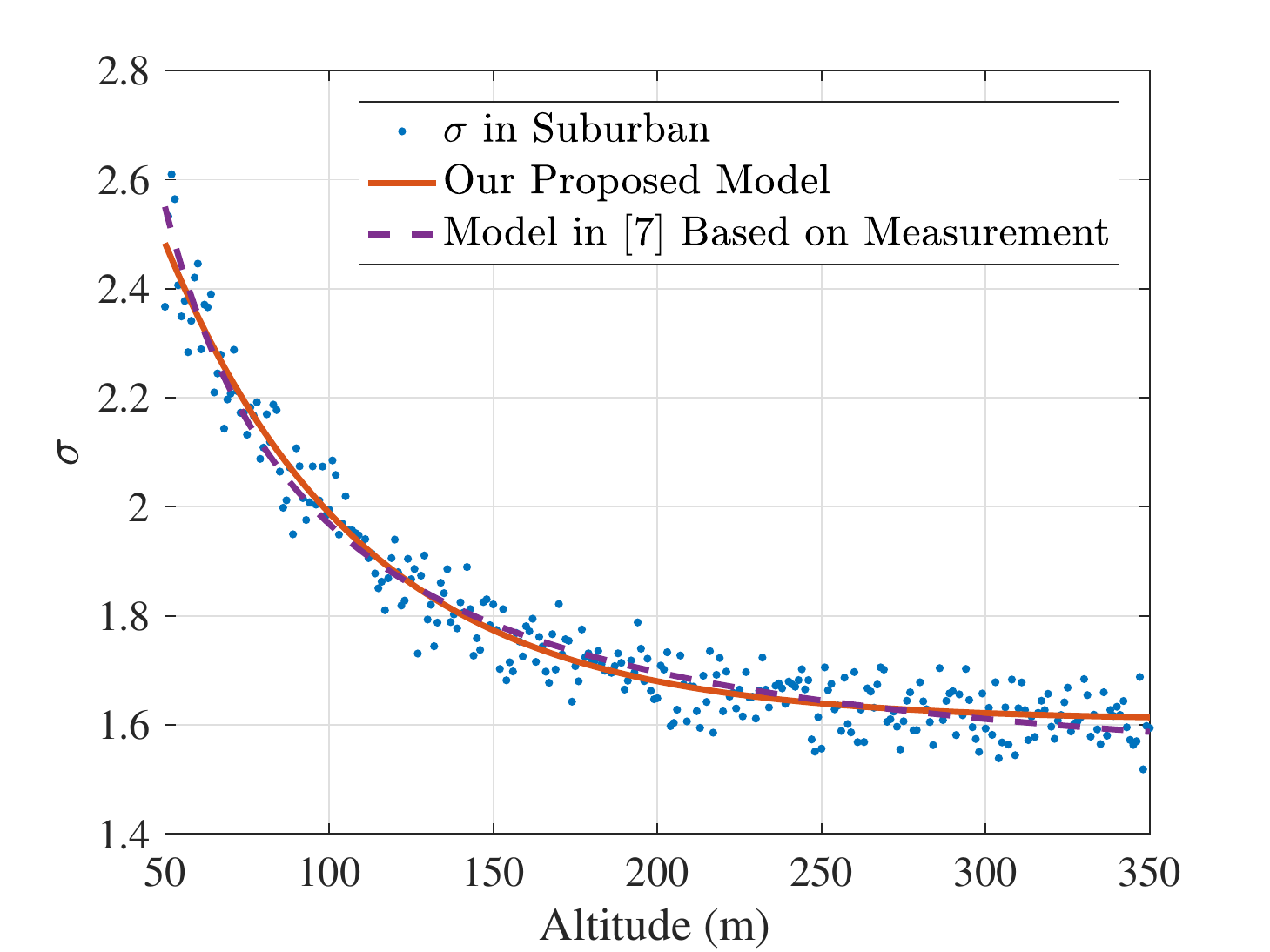}
\caption{Validation for our proposed shadowing model with measurement [7].}
\label{fig_sim}
\end{figure}
\begin{table}[htbp]
  \centering
  \caption{Results for Altitude-Dependent Shadowing Factor Model} \label{table3}
\begin{tabular}{|c|c|c|c|}
  \hline
\textbf{Environment} & \textbf{$p_2$} & \textbf{$q_2$} & \textbf{$r_2$}\\
    \hline
Suburban & 2.013 & 0.0167 & 1.608\\
\hline
Urban & 1.002 & 0.0250 & 1.369 \\
    \hline
Dense Urban & 3.936  & 0.0286 & 1.405 \\
\hline
High-rise Urban  & 11.001  & 0.0222 & 1.286 \\
  \hline
\end{tabular}
\end{table}


\section{Conclusion}
In this paper, we have proposed an analytical model for predicting the path loss and shadow fading for AA channel in built-up environments. The ground reflection in AA channel is decomposed into two direct paths in AG channel so that the LOS probability is reasonably used to calculate the ground reflection probability. In addition, the proposed path loss model and the corresponding shadowing are validated by ray-tracing simulations. Furthermore, the altitude-dependent shadowing model agrees well with the measurement-based model. Therefore, our research could be useful for the system design and analysis of HAP-enabling wireless communication.

\section*{Acknowledgment}
This work was supported by the National Key R\&D Program of China under grant No. 2016YFB1200102-04.


\ifCLASSOPTIONcaptionsoff
  \newpage
\fi

\end{document}